\documentclass[pdflatex,sn-mathphys-num]{sn-jnl}

\usepackage{graphicx}%
\usepackage{multirow}%
\usepackage{amsmath,amssymb,amsfonts}%
\usepackage{amsthm}%
\usepackage{mathrsfs}%
\usepackage[title]{appendix}%
\usepackage{xcolor}%
\usepackage{textcomp}%
\usepackage{manyfoot}%
\usepackage{booktabs}%
\usepackage{algorithm}%
\usepackage{algorithmicx}%
\usepackage{algpseudocode}%
\usepackage{listings}%

\theoremstyle{thmstyleone}%
\theoremstyle{thmstyletwo}%

\theoremstyle{thmstylethree}%

\begin{document}

\title[Article Title]{Reinforcement Learning-Guided Graph Transformations for SpTRSV Optimization}


\author*[1]{\fnm{Buse} \sur{Y{\i}lmaz}}\email{yilmazbuse@mef.edu.tr}

\affil*[1]{\orgdiv{Department of Computer Engineering}, \orgname{MEF University}, \orgaddress{\street{Huzur, Maslak Ayaza\u{g}a Cd.}, \city{\.{I}stanbul}, \country{Turkey}}}




\abstract{Sparse triangular solve (SpTRSV) is a fundamental kernel in numerous scientific and engineering applications. However, the data dependencies inherent in sparse triangular matrices significantly limit the available parallelism and make efficient workload distribution challenging. Recent graph transformation techniques address these limitations by modifying the dependency graph of the input matrix to improve parallel execution. Existing graph transformation strategies, however, rely on manually designed heuristics, making their development and adaptation to different optimization objectives challenging. This work proposes a reinforcement learning-guided graph transformation framework for SpTRSV, in which graph transformation is formulated as a sequential decision-making problem and an RL agent learns matrix-dependent transformation policies. Experimental results on real-world sparse matrices demonstrate level reductions of up to $94\%$ and reductions of up to $80\%$ in the coefficient of variation of level costs, while modifying only $1.50\%$ of the rows in the highest case. On average, the RL-guided graph transformation achieves a $23\%$ reduction in the number of levels and a $29\%$ reduction in the coefficient of variation of level costs while rewriting only $0.82\%$ of the matrix rows. Although the heuristic strategies generally achieve more aggressive level reduction(between $31\%$ and $46\%$), the RL-based approach achieves the largest average reduction in the coefficient of variation of level costs, demonstrating its ability to balance competing graph transformation objectives. The results further show that the learned policies can be transferred to previously unseen matrices through curriculum learning and fine-tuning, while zero-shot experiments provide insights into the limitations of generalizing graph transformation policies across different sparsity patterns.}

\keywords{Sparse triangular solve, reinforcement learning, parallel computing, performance optimization, graph transformation}



\maketitle

\section{Introduction}\label{section:intro}
There are several different approaches to optimize the parallel execution of Sparse Triangular Solve (SpTRSV) in the literature. Block diagonal approaches \cite{CUGU2020371,bradley_2016,mayer2009,totoni_2014_structure_adaptive,ahmad_2021_split_execution} partition the graph while keeping the sparsity pattern fixed and map these partitions to appropriate architectures and  SpTRSV approaches. Hence, these methods aim to optimize the execution mapping and its scheduling while the equation rewriting method \cite{byilmaz_basarim2020} modifies the sparsity pattern to optimize the schedule itself. Graph transformation strategies  \cite{byilmaz_efficient_graph_trans_strategies_9eylul2026, byilmaz_ccpe2022} leverage this method to apply structure-aware schedule optimization of SpTRSV by transforming the directed acyclic graph (DAG) of a sparse matrix into a more homogeneous and balanced graph by heuristic approaches. These strategies reduce the synchronization depth (critical path of the DAG) and improve workload balance to optimize the parallel execution efficiency of SpTRSV. In this sense, graph transformation as an optimization approach can be considered close to compiler optimizations such as IR transformations or loop transformations rather than classical SpTRSV tuning. Graph transformation problem \cite{byilmaz_basarim2020} presents a large combinatorial search space, as every rewriting operation changes the graph structure and consequently affects future rewriting opportunities. 

In this work, rather than designing increasingly complex hand-crafted heuristics, we formulate graph transformation as a reinforcement learning problem. The goal is not only to obtain high-quality graph transformations but also to provide a framework for systematically observing how local rewriting decisions affect the global structure of the DAG. This enables the learned policy to serve both as an optimization method and as a tool for analyzing the effectiveness of different graph characteristics and transformation criteria, thereby facilitating the development of future heuristic or learning based strategies.

Using the proposed framework, a reinforcement learning (RL) based graph transformation strategy is developed to optimize SpTRSV.  A subtle difference of this RL based strategy from the graph transformation strategies in the literature \cite{byilmaz_efficient_graph_trans_strategies_9eylul2026, byilmaz_ccpe2022} is providing a structure-aware learning policy rather than relying on the hand-crafted heuristics based on sparsity pattern features. The heuristic-based strategies prescribe the desired graph structure, whereas the RL agent learns a sequence of graph transformations that maximizes a long-term objective. Therefore, two approaches are optimizing at different levels of abstraction.  


Our contributions are summarized below:
\begin{itemize}
\item We develop an extensible reinforcement learning environment that serves as an automated heuristic discovery framework for graph transformation, enabling systematic evaluation of both learned policies and heuristic-based strategies.
\item We propose a novel graph transformation strategy for SpTRSV by formulating graph transformation as a reinforcement learning problem, replacing manually designed transformation rules with a learned policy. 
\item We demonstrate that reinforcement learning can effectively learn graph transformation policies that optimize multiple competing objectives, reducing the need for manual heuristic engineering.
\item We design a multi-objective reward function balancing critical-path reduction, level workload balance, rewriting cost and transformation progress.
\item We evaluate the learned transformation policies on real-world sparse matrices and demonstrate level reductions of up to $94\%$ and up to $80\%$ reduction in the coefficient of variation of level workloads, while requiring graph modifications to only a small fraction of the rows which goes as high as $1.58\%$.  The results demonstrate RL-based graph transformation's ability to balance competing graph transformation objectives. On average, the RL-guided graph transformation achieves a $23\%$ reduction in the number of levels and a $29\%$ reduction in the coefficient of variation of level costs while rewriting only $0.82\%$ of the matrix rows. The heuristic-based strategies \cite{byilmaz_efficient_graph_trans_strategies_9eylul2026} achieve average level reduction between $31\%$ and $46\%$ and $19\%$ to $22\%$ reduction in the coefficient of variation while rewriting a percentage of rows between $0.33\%$ and $0.81\%$. 
\end{itemize}

The organization of the paper is as follows: Section \ref{section:background} provides the fundamentals of SpTRSV, equation rewriting method and reinforcement learning, Section \ref{section:motivation} introduces the problem definition, Section \ref{section:methodology} explains the reinforcement learning based graph transformation framework in detail, Section  \ref{section:exp_res} presents the experiment results and Section \ref{section:conclusion} concludes while introducing the future work and the last section presents the literature work.

\section{Background} \label{section:background}
\subsection{Sparse Triangular Solve and Level-Set Method} \label{subsection:sptrsv}
The sparse triangular solve operation solves the linear system $Lx = b$, where $L$ is a sparse lower triangular matrix and $b$ is a dense right-hand side vector. Solving $Lx=b$ requires computing each unknown entry in $x$ vector only after all of its predecessors have been computed. These dependencies are represented as a directed acyclic graph (DAG), where each node corresponds to a row of the triangular system and each edge represents a computational dependency.

Sparse triangular solve is usually parallelized on CPUs using the level-set method  ~\cite{AndersonS89_Saad,Li2013,naumov_2011,rothberg_1992_parallel_iccg,Saltz:1990:aggregation}, which constructs a directed acyclic graph (DAG) representing the data dependencies among matrix rows and partitioning it into levels horizontally such that all rows within the same level are independent and can therefore be processed concurrently. During execution, all rows in a level are assigned to a group of threads and computed in parallel, while synchronization barriers enforce the dependencies between consecutive levels. As a result, the levels are processed sequentially despite the parallelism available within each level.

The effectiveness of the level-set method largely depends on the structural characteristics of the sparse matrix. Irregular sparsity patterns frequently produce levels with highly imbalanced computational workloads or only a few rows, therefore, a few computations. Levels with very few computations underutilize the available processing resources, whereas heavily populated levels may introduce additional communication and scheduling overhead when their workload is distributed among multiple threads. Furthermore, matrices with long dependency chains require many synchronization barriers, increasing execution overhead and limiting scalability.

\subsection{Equation Rewriting} \label{subsection:eq_rewriting}
To address the limitations and challenges posed by the level-set methods, the equation rewriting method~\cite{byilmaz_basarim2020} was introduced as a graph transformation technique for restructuring the dependency graph before execution. The method modifies the dependency structure by replacing selected row dependencies with their predecessors, thereby preserving correctness while enabling rows to migrate to earlier levels in the DAG. This transformation reduces the critical path by eliminating levels whenever possible and improves workload balance by merging computationally sparse levels. Consequently, the transformed DAG exposes greater parallelism while reducing synchronization overhead.

As indicated in \cite{byilmaz_ccpe2022,byilmaz_efficient_graph_trans_strategies_9eylul2026}, equation rewriting is performed entirely as a preprocessing step and is independent of the underlying SpTRSV solver. After transformation, the modified dependency graph is stored in CSR format and can be processed by existing level-set-based solvers that can matrices in CSR format without requiring any changes to the solver implementation. A detailed description of the equation rewriting methodology and its heuristic-based graph transformation strategies is available in ~\cite{byilmaz_basarim2020,byilmaz_ccpe2022,byilmaz_efficient_graph_trans_strategies_9eylul2026}.

\subsection{Reinforcement Learning}
Reinforcement learning (RL) \cite{sutton2018} is a machine learning paradigm in which an agent learns to make sequential decisions by interacting with an environment. At each time step, the agent observes the current state of the environment, selects an action according to a policy, and receives a reward that evaluates the quality of the selected action. The environment subsequently transitions to a new state, and the interaction continues until a terminal condition is reached. The objective of the agent is to learn a policy that maximizes the expected cumulative reward over an episode.

Unlike supervised learning\cite{sutton2018}, reinforcement learning does not require labeled training data. Instead, the agent discovers effective decision-making strategies through trial-and-error interactions with the environment. This characteristic makes RL particularly suitable for sequential optimization problems in which the quality of a decision depends not only on its immediate effect but also on its long-term consequences. Consequently, RL has been successfully applied to a wide range of optimization problems, including scheduling, resource allocation, compiler optimization, and graph-based learning tasks \cite{kaelbling1996,murphy2024rl}.

\subsection{Proximal Policy Optimization}
Proximal Policy Optimization (PPO)~\cite{schulman2017ppo} is a policy-gradient reinforcement learning algorithm that has gained widespread adoption due to its training stability and implementation simplicity. PPO employs an actor--critic architecture consisting of a policy network, which determines the probability of selecting an action, and a value network, which estimates the expected cumulative reward of a state. During training, the policy is updated using a clipped surrogate objective that restricts the magnitude of policy changes between successive updates to prevent excessively large parameter updates that may destabilize learning while maintaining efficient policy improvement.

Compared with earlier policy-gradient methods, PPO achieves a favorable balance between computational efficiency, robustness, and sample efficiency, making it well suited for environments with large and discrete action spaces. Owing to these advantages, PPO has become one of the most widely used reinforcement learning algorithms and serves as the learning algorithm adopted in this work.

\section{Motivation} \label{section:motivation}
Equation rewriting \cite{byilmaz_basarim2020} provides an effective mechanism for restructuring dependency graphs and existing approaches \cite{byilmaz_efficient_graph_trans_strategies_9eylul2026, byilmaz_ccpe2022}  rely on manually designed heuristics to determine which rows should be rewritten. Designing such heuristics requires significant domain expertise and often involves balancing multiple, sometimes conflicting, optimization objectives. Reinforcement learning offers an alternative by learning graph transformation policies directly through interaction with the rewriting environment.

Rather than focusing solely on the efficient execution of sparse triangular solve (SpTRSV), graph transformation techniques \cite{byilmaz_efficient_graph_trans_strategies_9eylul2026, byilmaz_ccpe2022} aim to optimize the dependency graph itself before execution. By determining which rows are rewritten, which dependencies are removed or introduced, and how the resulting levels are formed, graph transformation directly influences the execution schedule of the level-set method. The quality of the transformed graph determines important execution characteristics, including the critical path length, workload distribution across levels, synchronization frequency, and ultimately the achievable parallel performance.

Designing effective graph transformation strategies is challenging because the optimization objectives are strongly interdependent and often conflicting. For example, collapsing levels shortens the critical path but may significantly increase the computational cost of the remaining levels. Likewise, improving workload balance may reduce opportunities for future level collapses, while transformations that reduce synchronization overhead may adversely affect memory locality or communication costs. Existing graph transformation approaches \cite{byilmaz_efficient_graph_trans_strategies_9eylul2026, byilmaz_ccpe2022} therefore rely on manually designed heuristics based on matrix features, such as the average level cost (ALC), average number of rows per level (ARL), and average number of incoming dependencies per row (AIR), to balance these competing objectives. Although effective, such model-driven heuristics require considerable domain expertise and are inherently limited by the assumptions embedded in their design.

In this work, graph transformation is formulated as a reinforcement learning problem to automate heuristic discovery and design new graph transformation strategies. Instead of manually defining fixed rewriting rules, a reinforcement learning agent learns graph transformation policies through interaction with the graph rewriting environment and optimization of a multi-objective reward function. This enables the exploration of graph transformations that may not be considered by manually designed heuristics and provides a flexible policy-driven framework in which alternative optimization objectives, observation features, and reward formulations can be investigated without redesigning the underlying graph rewriting algorithm. Unlike previous heuristic-based strategies, which primarily rewrite rows between levels with a  few rows only, the proposed framework allows the learned policy to select source levels from the entire graph while suitable target levels are chosen primarily from  levels with a few rows only and from the rest as a fallback strategy. This additional flexibility enables the discovery of alternative graph topologies that may provide improved workload balance and shorter critical paths.

The proposed framework builds upon the same deterministic graph rewriting engine used by existing heuristic-based approaches. Reinforcement learning is responsible only for determining the sequence of graph transformations, whereas the graph rewriting engine performs the actual dependency modifications while preserving correctness. Consequently, the proposed approach does not replace graph rewriting; instead, it replaces manually designed heuristic decision rules with a learned graph transformation policy. Therefore, the proposed framework provides an extensible platform for automatically discovering execution-aware graph transformation strategies.
 
 \section{Methodology} \label{section:methodology}
 \begin{figure}[h]
 	\centering
 	\includegraphics[width=0.9\textwidth]{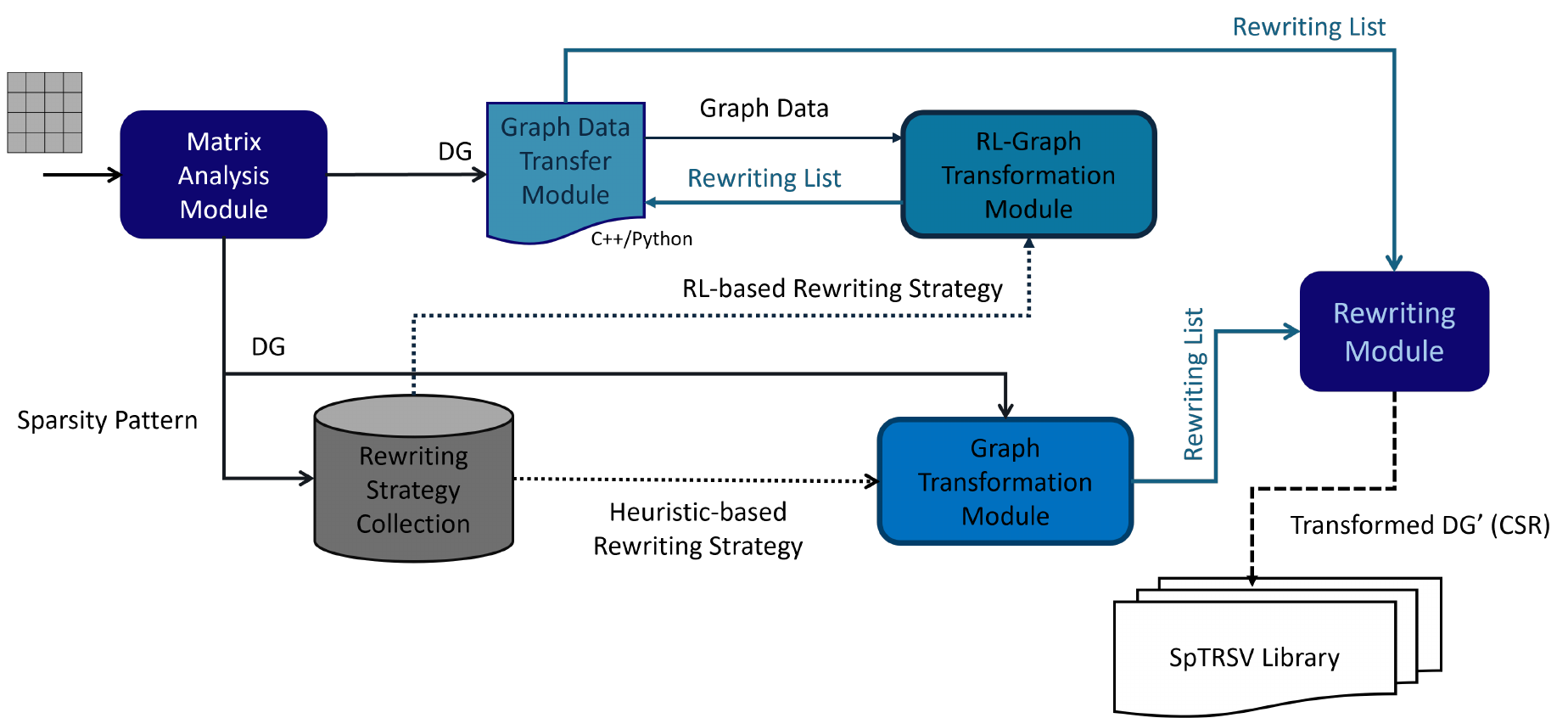}
 	\caption{Graph Transformation Framework Chainbreaker and the relationship between its modules. DG: dependency graph.}\label{fig:framework}
 \end{figure}
 Figure~\ref{fig:framework} illustrates Chainbreaker framework. The Reinforcement Learning Graph Transformation (RLGT) module and the Graph Data Transfer (GDT) module are introduced in this work  as extensions to the graph transformation framework presented in \cite{byilmaz_ccpe2022, byilmaz_efficient_graph_trans_strategies_9eylul2026}.  The RLGT module formulates graph transformation as a reinforcement learning problem.  We refer to the whole framework as Chainbreaker to differentiate between the main framework and the RL-based framework (RLGT and GDT modules) introduced in this work. The GDT module is implemented in C++ using pybind11 as the communication layer between the C++ framework and the Python-based RLGT module. It serves as the interface between the reinforcement learning components and the remaining modules of the framework. Since the overall architecture has been described in \cite{byilmaz_ccpe2022, byilmaz_efficient_graph_trans_strategies_9eylul2026} previously, only the components relevant to the reinforcement learning extension are presented in this section.
 
When the reinforcement learning graph transformation strategy is selected, the framework operates in either training or inference mode and GDT module prepares the DAG together with the level table containing the rows on each level and level costs. The prepared data object is transferred to RLGT Module where a Maskable PPO is used to learn a graph transformation policy. RLGT module generates a rewriting list just as its heuristic-based counterpart. The generated rewriting list is sent back to GDT Module and from there to the rewriting module which applies the actual graph transformation on the DAG. After the graph transformation is completed, the transformed DAG is converted back into a lower triangular matrix and stored on disk in CSR format by the write module. 
 
The following subsections describe the formulation of the environment, including the observation and action spaces, target selection strategy, reward function, and training procedure.

 \subsection{Environment} \label{subsection:environment}
 The environment models the current state of the dependency graph throughout the graph transformation process. At the beginning of an episode, the environment is initialized from the graph data received from GDT Module. Besides the graph structure, two matrix feature-based criteria that are also used in heuristic-based graph transformation strategies \cite{byilmaz_efficient_graph_trans_strategies_9eylul2026}, namely, Average Level Cost (ALC) and  Average Row Length (ARL) are computed and remain constant throughout the episode.
In \cite{byilmaz_efficient_graph_trans_strategies_9eylul2026}, a level is classified as thin if its computational cost is smaller than the average level cost (ALC) and thick otherwise. Thin levels represent candidate source and target levels. Removing nodes from thin levels can eventually empty them, hence the critical path of the graph can be shortened by collapsing the empty levels. In addition, moving nodes into target levels can improve workload balance across levels without immediately overloading the level since thin target levels become thicker.
 
 The original heuristic defines thin levels based on ALC, however this definition becomes problematic for the RL-based graph transformation strategy due to high AIR values or outlier coeficient of variation (CV) of level cost values. Matrices with high AIR may trigger large increases in level costs after rewriting, causing thin levels to disappear rapidly and reducing the number of feasible target levels. As a result, the agent's exploration space shrinks and episodes may terminate prematurely.
 
 In addition, when a matrix has a low CV of level cost, most level costs are close to the average. Consequently, after several rewrites, these levels quickly become thick, causing the agent's exploration space to shrink. This effect is amplified for matrices with high AIR values. Conversely, defining thin levels solely based on ALC becomes insufficient for matrices with a high CV of level cost, where the level cost distribution is highly skewed. ALC alone does not capture whether level costs are concentrated around the average or widely dispersed. As a result, the agent may prematurely exhaust its search space, causing episodes to terminate after only a small number of rewrites. On the other hand, when many thin target levels remain available, the agent may continue thickening them without inducing level collapses, resulting in unnecessarily long episodes before convergence.
 
 Under these conditions, the destination-selection heuristic becomes ill-conditioned because it relies on the availability of thin target levels. To preserve a sufficient exploration space throughout an episode, the RL environment incorporates two mechanisms: (1) defining the thin-level threshold as a function of ALC and CV of level costs, and (2) introducing a fallback mechanism for target level selection to avoid premature episode termination. These modifications preserve the original destination-selection heuristic whenever possible while preventing artificial episode termination.
 
 To preserve the exploration space under these conditions, we adapt the definition of the thin-level threshold as follows:
 
 $thin\_level\_threshold = ALC * (1.0 + alpha * max(0,1.0 - cv\_level\_cost))$
 
 The hyperparameter alpha controls the degree of relaxation. The term $(1 - CV)$ ensures that the multiplicative factor applied to ALC is maximized for low CV values, where a greater relaxation of thin\_level\_threshold is required. Since CV is not bounded by $1$, $max(0, 1 - CV)$ guarantees that the relaxation term becomes zero whenever $CV > 1$, preserving the original threshold.
 Consequently, the threshold is increased only for matrices with low CV, while matrices that do not exhibit this characteristic continue to use ALC value.
 
 After each rewriting operation, the environment incrementally updates the following values without modifying the DAG, allowing efficient interaction with the RL agent. 
 \begin{itemize}
 	\item node-to-level assignments, 
 	\item computational cost of each level, 
 	\item number of nodes per level, 
 	\item level indegrees, 
 	\item thin level set, 
 	\item total number of nonempty levels. 
 \end{itemize}
 
 \subsection{State Representation} \label{subsection:state}
 The observation contains all information required to determine the next environment state after applying a rewriting action. The environment state consists of $9$ global features together with $6$ level-specific features describing the current graph state. These features are provided in Table \ref{table:env_features}. All the values are kept in a normalized form.

 \begin{table}[h]
 	\caption{Global information about the current graph state and local information describing individual levels.}\label{table:env_features}%
 	\begin{tabular}{@{}ll@{}}
 		\toprule
 		Level-specific features & Global features \\ 
 		\midrule
 		Computational cost of each level & Current and reference CV of level costs \\ 
 		Number of nodes in each level & Current and reference CV of level node counts \\ 
 		Source level affinity values & Level count \\ 
 		Target level affinity values & Rewrite ratio \\ 
 		Completion ratio of each level & Thin levels completion ratio \\ 
 		Thin levels & Critical path ratio \\ 
 		~ & Thin-level ratio \\ 
 		\botrule
 	\end{tabular}
  \end{table}
 
 The agent selects one action (source level) from the masked valid actions. Therefore, at each step different source levels can be selected. Since reducing the critical path requires eliminating entire levels, the agent must repeatedly select the same source level until it becomes empty. Given the large exploration search space, relying only on the valid action mask is insufficient to achieve such goal. Therefore, each level is assigned source and target affinity values which are used to indicate the "hotness" of a level, guiding the agent to continue working on the same source level until it becomes empty while repeatedly selecting the same target level to create localized node accumulation on the graph.  At each step, the affinity values associated with the selected source and target levels are updated to reflect recent rewriting activity. The affinity scores gradually decay over time, allowing the agent to prioritize recently visited levels while avoiding permanent bias. Levels that become empty or cease to be suitable candidates receive negative affinity values to discourage their future selection.
 
 Completion ratio is used for a previous source level which is abandoned without being emptied. It is calculated as $1.0 - (curr\_level\_size / (init\_level\_size + 1))$. Based on this value, a certain penalty is calculated as part of the reward function. The observation space keeps completion ratio for each level as well as a global feature. The completion\_ratio\_thin is a  global feature that is the ratio of the current number of thin levels to the initial thin level count measuring how many thin levels are remaining as a working space for the agent. When this ratio falls below a certain threshold, the episode is terminated.
 
 Coefficient of variation of level costs and node counts per level are observed since one of the objectives is to balance the workload across levels. As the agent progresses, the number of nodes and the total cost of each level changes as well as the number of levels. These quantities evolve throughout the episode as graph transformations modify the workload distribution across levels. The reference coefficient of variation values are kept at $1.0$. 
 
 Rewrite ratio is the ratio of total number of rewrites to the total number of nodes. The agent might rewrite a previously rewritten row again as opposed to the heuristic-based strategies where a row is rewritten from a source level to a target level only once. Due to its sequential decision-making nature, the RL-based strategy transforms the graph incrementally by performing one rewrite at a time while exploring different graph configurations, rather than applying a single predetermined rewrite per row. Consequently, the RL-based strategy inherently incurs a higher rewriting overhead than its heuristic-based counterparts. Rewriting is a costly operation, therefore reducing the rewriting cost requires the maximum level reduction and coefficient of variation of level costs reduction with minimum number of rewrites. Critical path ratio observes the ratio of the current number of levels to the initial number of levels in order to track the level count reduction.
 
 \subsection{Action Space} \label{subsection:action}
 The action space consists of selecting a source level from which a row will be picked for rewriting, in other words, the node will be moved to an upper level. Although the rewriting happens at node level, the actions chosen by the agent are not nodes but levels. This design considerably reduces the action space compared to selecting individual nodes while still allowing the agent to influence the rewriting process. As a straightforward and simple approach, the node with the smallest number of dependencies is chosen to be rewritten aiming at minimizing the rewriting cost. 
 
 Not every level represents a legal action. A level is considered valid only if it satisfies the constraints defined by the rewriting algorithm: rewriting happens only to upper levels and between nonempty levels. For example, level 0 can only be a target level. Therefore, invalid actions are removed using Maskable PPO action masking. The policy network only samples from valid source levels, substantially reducing the effective search space.
 
 Two simple heuristics are used to further mask the actions. If a source level contains only one remaining node, all other actions are masked so that the agent is forced to complete the collapse of that level. The second heuristic is activated when a stagnant period is detected where there is no level collapse for a certain number of steps. The source affinity values are used to retain X source levels with the highest affinity as valid actions. X is a tunable parameter.
 
 \begin{algorithm}[t]
 	\caption{Move node to next thin level}
 	\label{alg:move_next_thin}
 	\begin{algorithmic}[1]
 		\Require Node $n$, source level $l_s$
 		\Ensure Target level $l_t$
 		
 		\State $l_t \gets$ \Call{FindPrevLevel}{$l_s$}
 		
 		\If{$l_t$ is thin}
 	    	\If{$l_t > 0$}
 		        \State $l_t \gets$ \Call{SelectHighestAffinityThinLevel}{$l_t$, SEARCH\_WINDOW}
 		    \Else
 		        \State $l_t \gets$ \Call{FindNearestNonEmptyLevel}{$l_s$}
 		        \If{$l_t$ is empty}
 		            return -1
 		        \EndIf
 		    \EndIf
 		\Else
  		   \If{$l_t > 0$}
 		      \State $l_t \gets$ \Call{SelectHighestAffinityThinLevel}{$l_t$, SEARCH\_WINDOW}
 		    \Else
 		      \State $l_t \gets$ \Call{FindNearestNonEmptyLevel}{$l_s$}
 		       \If{$l_t$ is empty or $l_t$ is thick}
 	 	          return -1
 		       \EndIf
 		    \EndIf
 		\EndIf 
 		
 		\If{$l_s - l_t > MAX\_REWRITE\_DISTANCE$}
 		   return -1
 		\EndIf
 		
 		\State \Call{UpdateGraphState}{$n, l_s, l_t$}
 		\State Append $\langle n, l_s, l_t \rangle$ to rewriting list
 		
 		\State \Return $l_t$
 		
 	\end{algorithmic}
 \end{algorithm}
 
 The RL agent selects the source level with the helps of action masking. Once a source level is selected, an appropriate target level is selected using the algorithm presented in Algorithm \ref{alg:move_next_thin}. Using the tunable parameter SEARCH\_WINDOW, the thin level with the highest target affinity among the X closest upper thin levels is selected as the target level. If there is no thin level above the source level, as a fallback strategy, the nearest nonempty upper level is chosen as the target level. If the selected source level is thick and there is no upper thin level, the action is considered invalid and the graph state remains unchanged. Furthermore, if the rewrite distance between the source and target levels exceeds a predefined threshold (MAX\_REWRITE\_DISTANCE), the action is also considered invalid and the graph state remains unchanged. The target level selected using the fallback strategy is typically a thick level. Moving a node from a thin level to a thick level is considered a legal action since one of the objectives is to collapse thin levels. Therefore, even if the target level is thick, this move is allowed. In contrast, rewriting from a thick source level to a thick target level is not allowed since it neither contributes to collapsing a level nor reduces the coefficient of variation of level costs. 
 
 Once the source and the target levels are chosen, the deterministic rewriting algorithm determines the new graph state, updates the graph state data structures without modifying the dependency graph and registers the rewritten row to the rewriting list together with the source and target levels.

 \subsection{Reward Function} \label{subsection:reward}
 The reward function is designed to encourage graph configurations that improve parallel SpTRSV  execution while avoiding transformations that degrade graph quality and preserving efficient graph transformations.
 Positive rewards are assigned when transformations reduce the number of levels(critical path of the graph) or improve level workload balance. Penalties are applied when the level costs becomes more imbalanced or when transformations fail to make useful progress.
 The reward combines multiple objectives are listed below:
 \begin{itemize}
 	\item reduction in the number of levels 
 	\item improvement in level cost balance
 	\item critical-path behavior 
 	\item movement efficiency 
 \end{itemize}
 
 This multi-objective reward allows the agent to evaluate long sequences of graph transformations instead of optimizing a single heuristic criterion. None of the reward components alone is sufficient to produce useful graph transformations; therefore they are optimized jointly.
 
 The reward consists of four components using coefficient of variation of level costs, critical path reduction, rewriting distance and abandonment penalty. The reward and penalty components are applied using a constant based on the initial level count and logarithmic functions are used to smooth the magnitudes. Level costs and critical path have the same weight while the abandonment penalty has a smaller weight. The rewriting distance has the smallest weight to avoid disrupting the target selection process. Since target-level selection is primarily determined by the affinity mechanism, rewriting distance is assigned a small weight so that it refines rather than dominates the policy.
 
 The calculation of the total reward and the weights of each reward component is provided below:
 
 $total\_reward = 0.80 * r\_cv\_level\_cost + 0.80 * r\_cpath + 0.05 * r\_rwdist + 0.50 * r\_abandon$
 
 The components of the reward function are explained in the following subsections and the reward calculation algorithm is given in Algorithm \ref{alg:reward}.
 
  \subsubsection{Critical path reduction (r\_cv\_level\_cost):} 
 Coefficient of variation of level costs is calculated and a penalty is applied if it is increased (load balance is worsened) and a reward if it is decreased (load balance is improved).
 
 \subsubsection{Critical path reduction (r\_cpath):} A reward is given if the current action caused the source level to be collapsed, otherwise a penalty is applied. The rewards or penalties are determined proportionally to the number of nodes of the source level. If level collapse did not happen but the source level is thin, still a small reward is provided to encourage the agent to work on thin levels, if the level is thick, a penalty is applied. 
 
  \subsubsection{Abandon penalty (r\_abandon):} In order to detect abandoned levels, the last source level seen as well as the time step that it was first touched are recorded and compared to the current source level selected. The penalty for abandoning the previous source level is applied to the current action. The size of the abandoned level is determined, and its completion ratio (how far it is from being empty) is calculated. Based on the time steps that passed since the first touch and the completion ratio, a penalty is calculated. In this way, the agent is encouraged to work on a source level frequently if not consecutively since it is punished for letting a touched level get "cold" over time. Since the exploration space is large, if not guided, the agent is free to select any level as source level which makes it difficult to learn the policy to collapse the levels in a meaningful way and the agent randomly wanders among the levels. However, to balance exploration and exploitation, the penalty is not applied if there are still levels with only $1$ node.
 
 If the abandoned source level was previously a target level, its initial size is set to its current size and first touched time step is set to zero to reset its parameters fitting its new role (being a source level). A previous target level should not be attempted at as a source level since a level that previously served as a target is expected to accumulate nodes rather than lose them. Hence, any attempt to rewrite a node from it will be a wasted effort and penalizing the agent for abandoning such a level would contradict the target-selection strategy. Therefore, the completion ratio is set to almost $1.0$ and no abandonment penalty is applied. Likewise, if the abandoned source level is now empty, no penalty is applied.
 
  \subsubsection{Rewriting distance (r\_rwdist):} Rewriting cost is a function of the sparsity pattern and the rewriting distance. The policy selects the row with the smallest number of parents to keep the rewriting cost small but this is not a guarantee since the agent does not know beforehand the connectivity between the node and its parents passed throughout the rewriting process, it is revealed as the node is actually being rewritten. Even though rewriting distance reward does not fully capture the entire rewriting cost, rewarding small rewriting distances and punishing large ones helps the agent to adjust the rewriting cost. The selection of a target level with the highest affinity among the closest five upper thin levels or the fallback mechanism choosing the nearest thick level does not invalidate this reward due to its small weight. The rewriting distances can be large based on the number of empty levels residing between the source and the target level. Regardless, the policy will tend to experience larger rewriting distances towards the episode ends since there will be empty levels scattered on the level sequence.
 
 \begin{algorithm}[t]
 	\caption{Reward Computation}
 	\label{alg:reward}
 	\begin{algorithmic}[1]
 		\Require Current and previous graph statistics
 		\Ensure Total reward
 		
 		\State $r_{cv} \gets 0$
 		\If{$\Delta CV < -\epsilon$}
 		\State $r_{cv} \gets -\alpha$
 		\ElsIf{$\Delta CV > \epsilon$}
 		\State $r_{cv} \gets L\_count\_init \cdot \Delta CV$
 		\EndIf
 		
 		\vspace{1mm}
 		
 		\State $r_{cp} \gets 0$
 		\State Compute $s \gets \log(1+\textit{source\_level\_size}+1)$
 		\If{critical path is reduced}
 		\State $r_{cp} \gets \alpha * (init\_level\_count / level\_count) * (1 + 1 / s)$
 		\ElsIf{source level is thin}
 		\State $r_{cp} \gets \alpha / s$
 		\Else
 		\State $r_{cp} \gets -1.0 * \alpha * (1 + s)$
 		\EndIf
 		
 		\vspace{1mm}
 		
 		\State $r_{abandon} \gets 0$
 		\If{previous source level is abandoned}
 		\If{abandoned source level was a target level}
 		\State $abandon\_penalty \gets \Call{ComputeSourceLevelSizeAndCompletionRatio}$
 		\State $r_{abandon} -= abandon\_penalty$
 		\State Reset abandoned source level size and completion ratio
 		\If{levels with 1 node exist and abandoned level is thin}
 		\State   $r_{abandon} \gets 0.0$
 		\EndIf
 		\EndIf
 		\EndIf
 		
 		\vspace{1mm}
 		
 		\State $r_{rwdist} \gets 0$
 	   \If{$r_{rwdist} < threshold$}
 	   \State $r_rwdist \gets log1p(max(0, threshold - rwdist)) / init\_level\_count$
 	   \Else
 	    \State $r_rwdist \gets -log1p(max(0, rwdist - threshold)) / init\_level\_count$
 	   \EndIf
 		
 		\vspace{1mm}
 		
 		\State
 		\Return
 		$0.80r_{cv}
 		+0.80r_{cp}
 		+0.05r_{dist}
 		+0.50r_{abandon}$
 		
 	\end{algorithmic}
 \end{algorithm}
 
 \subsection{Learning Algorithm} \label{subsection:LA}
 The policy is trained using Maskable Proximal Policy Optimization (Maskable PPO), which extends PPO by supporting invalid action masking.
 
 During training, each episode begins from the original dependency graph. The agent repeatedly selects valid source levels, then the target level is selected using two heuristics. Although graph rewriting is performed at the node level, the agent operates on source levels, which reduces the exploration space and simplifies the action space. The action is finalized by the deterministic graph rewriting algorithm. The reinforcement learning strategy does not modify the graph directly. Instead, it constructs a rewriting list consisting of the rewritten node together with its source and target levels. This is the same approach used with every graph transformation strategy in the Chainbreaker framework. The rewriting list consists of triplets of the node, source level and target level. Training continues until a terminal condition is reached when either of the following two conditions are met: (1) thin level ration falls below $0.02$,  (2) for sufficiently large graphs, the episode reaches a predefined maximum number of steps, in which case the episode is truncated. When the training or the inference is finalized, the rewriting list is returned to the main framework where it is passed to the rewriting module.
 
 A curriculum learning approach is taken where the policy is trained using three relatively small matrices with different sparsity patterns. The resulting policy is then used during inference to guide graph transformations on unseen matrices. During inference, multiple independent runs are executed for each matrix, and the graph configuration producing the smallest coefficient variation of level costs and number of levels is selected as the final transformed graph. In addition, transfer learning approach is taken for matrices that are relatively larger than the ones used for training, aiming at model performance.

\section{Experiment Results} \label{section:exp_res}
The experiments are conducted using a subset of the real world matrices \cite{suiteSparse} from \cite{byilmaz_efficient_graph_trans_strategies_9eylul2026} on an $8$ core 11th Gen Intel \textregistered  Core \texttrademark i7-11800H machine with $2.30$GHz,16 GB of RAM and an RTX 3050\texttrademark GPU. The selected matrices exhibit diverse sparsity patterns, matrix sizes and level set characteristics. Table \ref{table:matrices} gives the kind of the problem, the number of rows, number of nonzero elements for the lower triangle (L) matrix and the number of levels.

\begin{table}[h]
	\caption{Matrices from SuiteSparse Matrix Collection ~\cite{suiteSparse} that are used in experiments. The number of nonzeros (\# of NNZ) are for the lower triangular part of the matrix (L).}
	\label{table:matrices}
	\begin{tabular}{@{}lllll@{}}
		\toprule
				\textbf{Matrix Name} & \textbf{Kind}  & \textbf{\# of rows} & \textbf{\# of NNZ (L)} & \textbf{\# of levels}  \\ 
				\midrule
				\textbf{bcsstk17}    & Structural Prob. & 10,974  & 219,812  & 1,332  \\ 
				\textbf{bcsstk37}    & Structural Prob. & 25,503  & 58,324    & 5,030  \\
				\textbf{cfd2}        & Comp. Fluid Dynamics Prob. & 123,440  & 2,605,669 &  4,357  \\ 
				\textbf{gearbox}     & 	Structural Prob. & 153,746 & 4,617,075 & 4,586  \\ 
				\textbf{lung2}       & Comp. Fluid Dynamics Prob. & 109,460  & 273,647   & 479 \\ 
				\textbf{PR02R}       & Comp. Fluid Dynamics Prob. & 161,070 & 4,174,236 & 2,838  \\
				\textbf{torso2}      & 2D/3D Problem & 115,967 & 574,718   & 513  \\
				\textbf{venkat01}    & Comp. Fluid Dynamics Prob. Seq. & 62,424  & 890,108  &  4,176  \\
				\botrule
		\end{tabular}
	\end{table}

The experiments are conducted using the ChainBreaker framework, extended with the proposed RLGT and GDT modules \footnote{https://github.com/harubyy/chainbreaker\_graph\_transformation.git}. Data exchange between the main framework and  and the RLGT module is performed through a pybind11 interface implemented on GDT module. On the RLGT module, train and run python modules are written to perform the experiments. 

The implementation is done in C++ and Python. Gcc version 15.2.1 is used with optimization level \texttt{-O3}, and Python version used is 3.14.3. Training and inference both are performed on the CUDA device in a python environment. Stable-baselines3 \cite{raffin2021stable} is used with a Maskable PPO and SubprocVecEnv. The \texttt{MlpPolicy} is trained using a learning rate of $0.0001$ for $30$ PPO rollout iterations with a rollout length of $1024$ timesteps. To limit training time for matrices with a large number of levels, episodes are truncated when the number of elapsed timesteps exceeds a predefined maximum episode length. The maximum episode length is determined from the initial number of levels of the input matrix and is defined as $\max(2 \times \textit{num\_levels}, 3 \times \textit{log\_checkpoint})$, where \textit{log\_checkpoint} denotes the logging frequency and is set to $2048$ timesteps.

The RL agent is trained using SubprocVecEnv with four parallel worker environments. Parallel environment execution increases sample collection throughput by allowing multiple episodes to be processed simultaneously. The total number of training timesteps remains unchanged, ensuring that the parallel implementation affects only training efficiency and not the experimental protocol or learning objective.

The cost introduced by the graph transformation process should stay in acceptable limits. Given that iterative solvers typically take a few hundreds of iterations \cite{byilmaz_autotuning2016}, heuristic-based graph transformation strategies \cite{byilmaz_efficient_graph_trans_strategies_9eylul2026} can provide fast transformations when compared a reinforcement learning based  graph transformation strategy. Since RL-based strategy is implemented in Python while heuristic strategies are implemented directly in C++, direct execution time comparisons are not meaningful. Accordingly, training cost is treated as an offline optimization cost, whereas inference time is reported separately. The inference execution times are the average of $5$ runs where the rewriting list leading to the best graph configuration out of these $5$ runs is chosen to be sent back to the main framework. Therefore, the inference is not deterministic. A logging mechanism is used to log several parameters tracking the environment that are logged at every checkpoint, at the end of full and truncated episodes. Some of the parameters are \textit{time step, total reward, average reward, number of levels, CV of level costs, thin level ratio, thin move ratio, fallback ratio, invalid target ratio} and the \textit{type of logging}.  

During training, the framework records environment statistics, episode summaries, level costs, level sizes and rewriting lists for every completed episode. These logs are subsequently used to analyze the graph transformations produced by the learned policies.

Since all graph transformation strategies in ChainBreaker - including the proposed RL-guided strategy - derive decisions from structural properties of the input matrix, the learned policy is inherently matrix dependent. Therefore, a generalized trained model may not respond well to the different sparsity patterns. To see how well a trained model respond to sparsity patterns with different characteristics  three categories of experiments are conducted:
\begin{enumerate}
   \item \textbf{Individual training:} Independent models are trained and evaluated on \textit{lung2}, \textit{torso2}, and \textit{bcsstk17}, which have relatively small initial numbers of levels but exhibit different coefficients of variation (CV) of level costs.
	\item \textbf{Transfer learning (Curriculum learning + Fine-tuning):} A curriculum model is first trained sequentially on \textit{lung2}, \textit{torso2}, and \textit{bcsstk17}. This pretrained model is then fine-tuned independently on each of \textit{bcsstk37}, \textit{cfd2}, \textit{gearbox}, \textit{PR02R}, and \textit{venkat01}, producing a specialized model for each matrix. Then the fine-tuned models are  evaluated on their respective matrices. 
	\item \textbf{Zero-shot inference:} The curriculum model is directly evaluated on on \textit{bcsstk37}, \textit{cfd2}, \textit{gearbox}, \textit{PR02R}, and \textit{venkat01} without any additional training.
\end{enumerate}

The following subsections evaluate these experiment results in terms of PPO learning behavior, graph transformation results and inference behavior.

\subsection{PPO Learning Behavior} \label{subsection:ppo_results}
\begin{table}[]
	\caption{PPO learning behavior for individually trained models, curriculum model trained using lung2, bcsstk17 and torso2, and fine-tuned models that use the curriculum model.}
	\label{table:ppo}
	\begin{tabular}{@{}llllll@{}}
		\toprule
	  & \multicolumn{5}{c}{\textbf{Individual Model}}   \\
	  	  \midrule
	    & \textbf{explained var.}    & \textbf{explained var.}    & \textbf{approx\_kl}        & \textbf{entropy\_loss}     & \textbf{entropy\_loss}     \\
	  & \multicolumn{1}{c}{(avg.)} & \multicolumn{1}{c}{(max.)} & \multicolumn{1}{c}{(avg.)} & \multicolumn{1}{c}{(min.)} & \multicolumn{1}{c}{(avg.)} \\
		\midrule
		lung2    & 0.127   & 0.432   & 0.007   & -3.890    & -3.467  \\
		bcsstk17 & 0.167    & 0.470    & 0.015   & -6.610  & -5.574   \\
		torso2   & -8.169E-05 & 1.980E-04 & 0.002    & -5.820   & -5.090   \\
		\midrule
		& \multicolumn{5}{c}{\textbf{Curriculum Model}}    \\
		\midrule 
		& \textbf{explained var.}    & \textbf{explained var.}    & \textbf{approx\_kl}        & \textbf{entropy\_loss}     & \textbf{entropy\_loss}     \\
		& \multicolumn{1}{c}{(avg.)} & \multicolumn{1}{c}{(max.)} & \multicolumn{1}{c}{(avg.)} & \multicolumn{1}{c}{(min.)} & \multicolumn{1}{c}{(avg.)} \\
		 \midrule
		lung2    & \multicolumn{1}{c}{-}   & \multicolumn{1}{c}{-}    & \multicolumn{1}{c}{-}    & \multicolumn{1}{c}{-}   & \multicolumn{1}{c}{-}       \\
		bcsstk17 & 0.353    & 0.802  & 0.014  & -6.750   & -5.643  \\
		torso2   & 0.107 & 0.503  & 0.014 & -5.53  & -4.720  \\
		\midrule
		& \multicolumn{5}{c}{\textbf{Transfer Learning (Curriculum learning + Fine Tuning)}}  \\
		\midrule
		& \textbf{explained var.}    & \textbf{explained var.}    & \textbf{approx\_kl}        & \textbf{entropy\_loss}     & \textbf{entropy\_loss}     \\
		& \multicolumn{1}{c}{(avg.)} & \multicolumn{1}{c}{(max.)} & \multicolumn{1}{c}{(avg.)} & \multicolumn{1}{c}{(min.)} & \multicolumn{1}{c}{(avg.)} \\
		\midrule
		bcsstk37 & 0.056   & 0.380  & 0.013 & -7.420  & -6.261  \\
		cfd2     & 0.000  & 0.000  & 0.005   & -8.180  & -8.153  \\
		gearbox  & 0.004  & 0.025 & 0.014  & -7.840  & -7.647  \\
		PR02R    & 0.083  & 0.189  & 0.017  & -7.730  & -7.686  \\
		venkat01 & 0.000  & 0.000 & 0.024  & -8.120    & -8.019   \\       
		 \botrule   
	\end{tabular}
\end{table}

The PPO learning behavior is evaluated in Table \ref{table:ppo} using explained variance, approximate KL divergence, and entropy loss. For the individually trained models, the average explained variance ranges from approximately $0$ to $0.167$, with maximum values of $0.432$ and $0.470$, respectively. The curriculum model is trained sequentially on \textit{lung2}, \textit{bcsstk17}, and \textit{torso2}; therefore, no separate curriculum-training statistics are reported for \textit{lung2}. For \textit{bcsstk17} and \textit{torso2}, curriculum learning increases the average explained variance to $0.353$ and $0.107$, respectively, with maximum values of $0.802$ and $0.503$. This indicates that knowledge accumulated from previously trained matrices improves value-function learning for subsequent matrices. The approximate KL divergence remains relatively small throughout the experiments, ranging from $0.002$ to $0.024$, indicating stable policy updates without excessive divergence between consecutive policies. The decrease in entropy loss observed with curriculum learning also indicates that the policy becomes more focused as training progresses while retaining stochasticity for exploration.

In contrast, the average explained variance obtained during transfer learning is substantially lower for the target matrices, ranging from $0.000$ to $0.083$, despite the stable approximate KL divergence. This behavior indicates that the value function learned from previously encountered matrices does not generalize well to different matrix structures. Such behavior is consistent with the nature of graph transformation for sparse triangular systems. Each sparse matrix has a distinct sparsity pattern, which determines the dependency structure and consequently the possible effects of rewriting operations. In addition, rewriting a row may decrease, preserve, or increase its indegree count, and therefore its computational cost, resulting in varying effects on the workload distribution across levels. Consequently, the effect of a graph transformation is highly dependent on the underlying sparsity pattern. These observations support the specialization of the graph transformation policy to the characteristics of the target matrix rather than assuming direct generalization across different sparse matrices. The transfer-learning results therefore motivate fine-tuning as a means of adapting the learned policy to the specific sparsity pattern and transformation behavior of each target matrix.

The results also motivate evaluating the learned policy in a zero-shot setting, in which the target matrices are not used for either curriculum training or fine-tuning. Since graph transformation behavior is strongly dependent on the underlying sparsity pattern, strong zero-shot generalization is not expected. Nevertheless, the experimental results presented in Sections~\ref{subsection:graph_tr_results} and~\ref{subsection:inference_behavior_results} indicate that zero-shot inference can achieve performance comparable to transfer-learning inference, and in some cases may even approach it. Zero-shot inference experiment is useful for quantifying the extent to which the learned policy captures transformation behavior that is independent of a specific matrix structure. 

\subsection{Graph Transformation Results} \label{subsection:graph_tr_results}
Table~\ref{table:graph_res} summarizes the graph characteristics before and after RL-based transformation for \textit{lung2}, \textit{bcsstk17}, and \textit{torso2}. The table reports initial characteristics such as number of levels, ALC, ARL, AIR values as well was CV of level costs before and after the RL transformation. In addition, for RL transformation, information about thin level cost threshold, number of thin levels and some other details are reported. The table also gives percentage of level reduction and train or inference runtime in seconds. The timings for inference are average of $5$ indeterministic runs. These matrices are used for both individual model training and the construction of the curriculum model. The initial characteristics show substantial differences among the matrices. In particular, \textit{lung2} has the highest initial coefficient of variation ($6.72$), whereas \textit{bcsstk17} has the largest number of levels ($1,332$). Although \textit{torso2} has a comparable number of levels to \textit{lung2}, its initial ALC and ARL are substantially higher making level reduction particularly difficult. These differences result in considerably different transformation outcomes across the matrices, demonstrating the matrix-dependent nature of the graph transformation problem.

The individually trained models reduce the number of levels by $94\%, 22\%, and 17\%$ for \textit{lung2}, \textit{bcsstk17}, and \textit{torso2}, respectively. The corresponding CV values are reduced by a value ranging between 43\% and 80\% during training. The results obtained during inference are highly consistent with those observed during training improving the level reduction counts except for \textit{lung2}. The final level counts are $30$, $1,015$, and $420$. The CV values are comparable for indiivdual training nad inference results. 

For the curriculum models, the results on \textit{bcsstk17} and \textit{torso2} remain comparable to those obtained with individually trained models. Since the model trained using \textit{lung2} for the curriculum model, the results are the same, hence they are not reported twice. The level reduction obtained with inference are higher for \textit{bcsstk17} and \textit{torso2} while remaining the same for \textit{lung2}.  The curriculum model is used for transfer learning for five other matrices. 

The results also highlight the different computational costs associated with training and inference. Individual model training requires approximately $81$s for \textit{lung2}, $79$min. for \textit{bcsstk17}, and $97$min. for \textit{torso2}, whereas the corresponding average inference times over five runs range between $6.25$s, and $319$s. Thus, once the policy has been trained, applying the learned transformation strategy is considerably less costly than training the model. The variation in both training and inference time across matrices is consistent with their substantially different graph sizes and structural characteristics.

\begin{table}[!htbp]
	\caption{Graph transformation results for training results for individual training, transfer learning and individual inference results.}
	\label{table:graph_res}
	\begin{tabular}{@{}llll@{}}
		\toprule
		\textbf{initial} & lung2 & bcsstk17 & torso2 \\
		\midrule
		num. of levels            & 479                       & 1,332                        & 513                        \\
		ALC                       & 914                       & 321                          & 2014                       \\
		ARL                       & 228.518                   & 8.23874                      & 226.057                    \\
		AIR                       & 1.49997                   & 19.0303                      & 3.95588                    \\
		CV of level cost          & 6.72  & 0.83     & 0.82   \\
		\midrule
		\textbf{RL transformation} & lung2 & bcsstk17 & torso2 \\
		\midrule
		threshold                  & 914.06                    & 432.2                        & 2370.64                    \\
		max node count   in thin   & 228                       & 64                           & 270                        \\
		max cost in   thin         & 912                       & 430                          & 2370                       \\
		initial thin   level count & 462                       & 920                          & 312                        \\
		\midrule
		\textbf{RL transformation} & lung2 & bcsstk17 & torso2 \\
		\textbf{}                  & \multicolumn{3}{c}{\textbf{individual training}}                                      \\
		\midrule
		num. of levels            & 29                        & 1039                         & 425                        \\
		ALC                       & 15151.448                 & 598.951                      & 2732.379                   \\
		ARL                       & 3774.483                  & 10.562                       & 272.864                    \\
		AIR                       & 1.507                     & 27.854                       & 4.507                      \\
		CV                        & 1.3298                    & 0.2922                       & 0.4742                     \\
		RL train time              & 81.32                     & 4727.34                      & 5792.57                    \\
		level reduction  ptg.          & 94\%  & 22\%     & 17\%   \\
		\midrule
		\textbf{RL transformation} & lung2 & bcsstk17 & torso2 \\
		\textbf{}                  & \multicolumn{3}{c}{\textbf{curriculum learning}}  \\
		\midrule
		num. of levels            & \multicolumn{1}{c}{-}     & 1040                         & 438                        \\
		ALC                       & \multicolumn{1}{c}{-}     & 594.362                      & 2614.902                   \\
		ARL                       & \multicolumn{1}{c}{-}     & 10.552                       & 264.765                    \\
		AIR                       & \multicolumn{1}{c}{-}     & 27.664                       & 4.438                      \\
		CV                        & \multicolumn{1}{c}{-}     & 0.2777                       & 0.5386                     \\
		RL train time              & \multicolumn{1}{c}{-}     & 8216.83                      & 4656.17                    \\
		level reduction ptg.           & \multicolumn{1}{c}{-}     & 22\%     & 15\%   \\
		\midrule
		\textbf{RL transformation} & lung2 & bcsstk17 & torso2 \\
		& \multicolumn{3}{c}{\textbf{individual inference}}                                   \\
		\midrule
		num. of levels            & 30                        & 1015                         & 420                        \\
		ALC                       & 14639.067                 & 595.494                      & 2771.302                   \\
		ARL                       & 3648.667                  & 10.812                       & 276.112                    \\
		AIR                       & 1.506                     & 27.039                       & 4.518                      \\
		CV                        & 1.368                     & 0.283                        & 0.465                      \\
		num. of rewritten rows	& 4,093	& 3,474	& 5,630  \\
		avg. RL run time & 6.252                     & 319.628                      & 79.101                     \\
		level reduction ptg.           & 94\%  & 24\%     & 18\%  \\
		\botrule
	\end{tabular}
\end{table}

\begin{table}[!htbp]
	\caption{Graph transformation results for training results using transfer learning and inference results for transfer learning and zero-shot experiments.}
	\label{table:graph_res2}
	\begin{tabular}{@{}llllll@{}}
		\toprule
		\textbf{initial}           & cfd2                    & gearbox                  & venkat01                & PR02R                   & bcsstk37   \\
		\midrule
		num. of levels            & 4,357                   & 4,586                    & 4,176                   & 2,838                   & 5,030           \\
		ALC                      & 708                     & 1,979                    & 411                     & 2,884                   & 226                      \\
		ARL                       & 28.33                   & 33.53                    & 14.95                   & 56.75                   & 5.07                     \\
		AIR                       & 12.01                   & 29.03                    & 13.26                   & 24.92                   & 21.87                    \\
		CV of level costs          & 0.57                    & 0.88                     & 0.46                    & 0.37                    & 0.76                     \\
		\midrule
		\textbf{RL transformation} & cfd2                    & gearbox                  & venkat01                & PR02R                   & bcsstk37      \\
		\midrule
		threshold                  & 1010.55                 & 2224.62                  & 635.26                  & 4710.3                  & 281.68       \\
		thin levels                & 3,485                   & 2,423                    & 3,685                   & 2,838                   & 3,732     \\
		max. node count in thin   & 44                      & 208                      & 27                      & 682                     & 11         \\
		max. cost in thin         & 1,010                   & 2,221                    & 635                     & 4,022                   & 281       \\
		initial thin level count & 3,485                   & 2,423                    & 3,685                   & 2,838                   & 3,732        \\
		\midrule
		\textbf{RL transformation} & cfd2                    & gearbox                  & venkat01                & PR02R                   & bcsstk37      \\
		\textbf{}                  & \multicolumn{5}{c}{\textbf{transfer learning   (curriculum learning + fine-tuning)}}     \\
		\midrule
		num. of levels            & 4250                    & 3953                     & 4037                    & 2809                    & 4093     \\
		ALC                       & 1371.726                & 2331.572                 & 470.816                 & 2975.421                & 394.324     \\
		ARL                       & 29.045                  & 38.893                   & 15.463                  & 57.341                  & 6.231     \\
		AIR                       & 23.114                  & 29.474                   & 14.724                  & 25.445                  & 31.143      \\
		CV                        & 0.554                   & 0.706                    & 0.467                   & 0.358                   & 0.496         \\
		RL train time              & 29128.97                & 119.10                   & 27809.72                & 59.88                   & 13299.36        \\
		total train \& I/O time & 29135.97                & 125.72                   & 27815.23                & 67.84                   & 13304.71 \\
		level reduction            & 2\%                     & 14\%                     & 3\%                     & 1\%                     & 19\%          \\
		\midrule
		\textbf{RL transformation} & cfd2                    & gearbox                  & venkat01                & PR02R                   & bcsstk37      \\
		\textbf{}                  & \multicolumn{5}{c}{\textbf{transfer learning (curriculum learning + fine-tuning)(inference)}}   \\
		\midrule
		num. of levels            & 4245                    & 3834                     & 3999                    & 2746                    & 3836        \\
		ALC                       & 1422.687                & 2406.347                 & 461.447                 & 3046.577                & 40.08      \\
		ARL                       & 29.079                  & 40.101                   & 15.61                   & 58.656                  & 29.497      \\
		AIR                       & 23.963                  & 29.504                   & 14.281                  & 25.47                   & 14.167         \\
		CV                        & 0.546                   & 0.673                    & 0.462                   & 0.329                   & 0.674        \\
		num. of rewritten rows	& 7763	& 9431	& 7726	& 5972	& 4667 \\
		avg. RL run time & 1.07E+04                & 33.796453                & 1.68E+03                & 23.949541      &       \\
		level reduction            & 3\% & 16\% & 4\% & 3\% & 24\% \\
		\midrule
		\textbf{RL transformation} & cfd2                    & gearbox                  & venkat01                & PR02R                   & bcsstk37      \\
		\textbf{}                  & \multicolumn{5}{c}{\textbf{curriculum learning (zero-shot)(inference)}}       \\
		\midrule
		num. of levels            & 4254                    & 3940                     &          3994               & 2815                    &     \\
		ALC                       & 1437.001                & 2347.215                 &         463.965                & 2968.474                &      \\
		ARL                       & 29.017                  & 39.022                   &            15.629          &  57.218                  &       \\
		AIR                       & 24.261                  & 29.576                   &        14.343       &  25.44                   &       \\
		CV                        & 0.518                   & 0.700                    &         0.4611         & 0.362                   &        \\
		num. of rewritten rows & 5998	& 4506	& 7786  & 	5976 & \\	
		avg. RL run time & 1.65E+04                & 29.407     &         436.516   & 25.021      &     \\
		level reduction            & 2\%                     & 14\%                     & 100\%                   & 1\%                     & 100\%     \\          
		\botrule    
	\end{tabular}
\end{table}

Table~\ref{table:graph_res2} uses the same structure as Table \ref{table:graph_res} and presents the transformation results for the five matrices used for transfer learning and compares transfer-learning inference with zero-shot inference. The matrices exhibit substantially different structural characteristics. Their initial number of levels ranges from $2,838$ to $5,030$, while the number of thin levels ranges from $2,423$ to $3,732$. The initial CV values range from $0.37$ to $0.88$, further demonstrating the variation in workload distribution among the matrices.

Transfer learning results in varying level reduction percentages across the target matrices which are increased during inference: the number of levels is reduced by a maximum value of 24\% for \textit{bcsstk37} and a minimum of 3\% for \textit{cfd2} and  \textit{PR02R}. The corresponding CV values are reduced by a value ranging between 4\% and 24\% while it remained the same for \textit{venkat01}. \textit{bcsstk37} demonstrates interesting results with an increase in the level reduction and a decrease in CV value while still being smaller than the original CV value.

The zero-shot results show that the curriculum-trained policy can produce meaningful transformations even without fine-tuning on the target matrix. The results indicate that the learned policy contains transformation behavior that can be applied to previously unseen matrices, although matrix-specific adaptation through fine-tuning provides additional improvement.
This observation is consistent with the matrix-specific nature of sparse graph transformations: although the learned policy captures reusable transformation behavior, the effect of individual rewrites depends on the sparsity pattern and the resulting changes in row dependencies and level workloads. Fine-tuning therefore allows the policy to exploit characteristics of the target matrix that cannot be fully captured through zero-shot generalization.

The results should also be interpreted with respect to the predefined inference step limit. Several transformations continue to reduce the level count and/or CV while approaching the available inference budget. Consequently, the reported values represent the transformation quality achieved within the prescribed computational budget and do not necessarily represent the maximum reduction that could be obtained through unrestricted inference. 

\subsection{Inference Behavior} \label{subsection:inference_behavior_results}

\begin{table}[!htbp]
	\caption{Inference behavior for individually trained models, the curriculum model and zero-shot learning}
	\label{table:inference}
	\begin{tabular}{@{}llllll@{}}
		\toprule
		\multicolumn{6}{c}{\textbf{Individual Model Inference}}   \\
		\textbf{}                       & \multicolumn{1}{c}{\textbf{lung2}}    & \multicolumn{1}{c}{\textbf{bcsstk17}} & \multicolumn{1}{c}{\textbf{torso2}}  &  &     \\
		\midrule
		\textbf{avg. reward}            & 1.115  & 0.043  & -1.076&  &     \\
		\textbf{cv level cost}          & 1.3684 & 0.2833 & 0.4647&  &     \\
		\textbf{levels}                 & 30  & 1015& 420&  &     \\
		\textbf{thin level   ratio}     & 0.3 & 0.018  & 0.319 &  &     \\
		\textbf{move count}             & 2502& 2619& 6144  &  &     \\
		\textbf{thin move   ratio}      & 0.746  & 0.691  & 0.726 &  &     \\
		\textbf{avg move dist}          & 31.354 & 19.846 & 33.035&  &     \\
		\textbf{fallback}               & 0   & 0.002  & 0  &  &     \\
		\textbf{invalid target count} & 0   & 1174& 0  &  &     \\
		\midrule
		\textbf{init. cv level cost}  & 6.72  & 0.83  & 0.82 & & \\
		\textbf{init. num. levels}& 479  & 1,332  & 513 &  & \\
		\textbf{reduction in num. levels}  & 94\%   & 24\% & 18\%  &  & \\
		\midrule
		\multicolumn{6}{c}{\textbf{Transfer Learning Inference}}  \\
		\textbf{} & \multicolumn{1}{c}{\textbf{bcsstk37}} & \multicolumn{1}{c}{\textbf{cfd2}}     & \multicolumn{1}{c}{\textbf{gearbox}} & \multicolumn{1}{c}{\textbf{PR02R}} & \multicolumn{1}{c}{\textbf{venkat01}} \\
		\midrule
		\textbf{avg. reward}            & 0.821  & -4.885 & 0.798 & 0.427   & -0.079 \\
		\textbf{cv level cost}          & 0.6737 & 0.5458 & 0.673 & 0.3286  & 0.4617 \\
		\textbf{levels}                 & 3836& 4245& 3834  & 2746& 3999\\
		\textbf{thin level   ratio}     & 0.41& 0.216  & 0.406 & 0.988   & 0.723  \\
		\textbf{move count}             & 5004& 7912& 5053  & 6144& 8352\\
		\textbf{thin move   ratio}      & 0.779  & 0.61& 0.776 & 0.985   & 0.909  \\
		\textbf{avg move dist}          & 14.167 & 52.949 & 14.474& 2.771   & 10.257 \\
		\textbf{fallback}               & 0   & 0   & 0  & 0& 0   \\
		\textbf{invalid target count} & 4168& 802 & 4119  & 0& 0  \\
		\midrule
		\textbf{init. cv level cost}  & 0.76  & 0.57  & 0.88 & 0.37  & 0.46  \\
		\textbf{init. num. levels}   & 5,030 & 4,357 & 4,586& 2,838 & 4,176 \\
		\textbf{reduction in num. levels}  & 24\%  & 3\% & 16\% & 3\%& 4\%  \\
		\midrule
		\multicolumn{6}{c}{\textbf{Zero-shot Inference}}\\
		& bcsstk37  & cfd2  & gearbox  & PR02R  & venkat01  \\
		\midrule
		avg\_reward&  & -5.26 & 0.394& 0.419  & -0.12 \\
		cv\_level\_cost &  & 0.5181& 0.6998   & 0.3624 & 0.4611\\
		levels &  & 4254  & 3940 & 2815   & 3994  \\
		thin\_level\_ratio  &  & 0.194 & 0.409& 0.992  & 0.71  \\
		move\_count&  & 7792  & 4774 & 6144   & 8352  \\
		thin\_move\_ratio   &  & 0.579 & 0.76 & 0.995  & 0.906 \\
		avg\_move\_dist &  & 56.576& 15.591   & 2.727  & 10.91 \\
		fallback   &  & 0& 0& 0  & 0\\
		invalid\_target count &  & 922   & 4398 & 0  & 0 \\
		\midrule
		\textbf{init cv level   cost}  & 0.76  & 0.57  & 0.88 & 0.37   & 0.46  \\
		\textbf{init num   levels}& 5,030 & 4,357 & 4,586& 2,838  & 4,176 \\
		\textbf{reduction in   num levels} &  & 2\%   & 14\% & 1\%& 4\%  \\
		\botrule
	\end{tabular}
\end{table}

Table~\ref{table:inference} presents the results collected from the logs that are obtained during inference for individually trained models, transfer-learning models and zero-shot learning. For each matrix, the best inference run among five independent runs is reported. Some of the information presented in this table such as reduction in the number of levels, CV of level costs, the number of levels and their corresponding initial values are already presented in Tables \ref{table:graph_res} and Table \ref{table:graph_res2}, and they are repeated here for the ease of interpretation.  
The table presents the resulting thin level ratio, total move count for reported duration (n $\times$ 2048 if not truncated/terminated),  move ratio where source is thin, average number of levels as the move distance, fallback ratio of total moves and invalid target count where the step is skipped and no rewriting happens. It is observed that the results are obtained primarily through moves originating from thin levels, with high thin-move ratios respectively with the minimum being $0.579$ for \textit{cfd2}, zero-shot inference. Fallback ratios are generalle zero, hence moves from thin to thick levels or from thick to either thin or thick levels do not happen. Although these cases are encountered, the move is rejected due to being from thick to thick level or distance exceeding the threshold. These values are indicated by the invalid target count.

The transfer-learning models also successfully transform the target matrices, although the amount of level reduction varies considerably. \textit{bcsstk37} achieves the largest reduction, from $5,030$ to $3,836$ levels ($24\%$), followed by \textit{gearbox}, which is reduced from $4,586$ to $3,834$ levels ($16\%$). The reductions for \textit{cfd2}, \textit{PR02R}, and \textit{venkat01} are more limited, at $3\%$, $3\%$, and $4\%$, respectively. The final CV values are reduced relative to the initial values for all transfer-learning cases except \textit{venkat01}, for which the initial and final values are approximately equal. The learned policies frequently select thin source levels, with thin-move ratios ranging from $0.61$ to $0.985$. As seen in Table \ref{table:graph_res2}, for \textit{PR02R}, due to thin level threshold being equal to the maximum level cost, the whole matrix is considered thin and maximum node count is $682$ and the CV of level costs is $0.37$. Given these characteristics, it is challenging to collapse levels, therefore the thin level ratio remains at $0.988$ and $0.992$ after the transfer learning and zero-shot inferences. A similar case for maximum node count is observed for \textit{gearbox} with $208$ nodes. However, a higher CV value of $0.88$ and lower initial thin level count of $2423$ compared to \textit{PR02R} help with the level reduction despite a lower thin move ratio of $0.776$. 


The transformation behavior also varies according to the distribution of level costs and the resulting changes in AIR. For \textit{cfd2}, AIR increases considerably during transformation, while thin levels can quickly become thick as their costs increase, making it more difficult for the agent to continue focusing on thin levels and limiting further level collapse. In contrast, \textit{gearbox} exhibits a clearer separation between thin and thick levels in terms of level costs allowing the agent to maintain its focus on thin levels, while AIR does not spike enough to become an obstacle to level reduction. A similar favorable separation between thin and thick levels is observed for \textit{bcsstk37}. The agent simply shuffled the nodes between levels due to AIR not increasing enough to saturate thin levels to become thick and the number of nodes in a level not decreasing fast enough to trigger level collapses. Another possibility is source levels becoming target levels frequently. The behavior of \textit{PR02R} follows a pattern similar to that observed for \textit{venkat01}. The level costs before and after the transformation for transfer learning inference are provided as graphs given in Figure \ref{fig:group1} and Figure \ref{fig:group2} in Appendix \ref{secA1}. When explained variance values from Table \ref{table:ppo} are considered, it is observed that the relationship between explained variance and transformation quality is not strictly one-to-one, still \textit{bcsstk37} has the highest level reduction and the highest explained variance while \textit{gearbox} and \textit{PR02R} follow next.


Several transfer-learning inference runs reach the predefined inference step limit, indicating that the policies continue to perform transformations until the available inference budget is exhausted. Therefore, the reported reductions represent the improvements obtained within the prescribed inference budget rather than necessarily the maximum reductions attainable by continued inference.Zero-shot inference results are similar to transfer learning inference results, zero-shot results for \textit{bcsstk37} are missing due to a bug. It is observed that the average rewards are usually around $0$ implying that more aggressive level collapsing approaches might improve the results. 


 
\begin{table}[!htbp]
\caption{Comparison of RL-based graph transformation strategy (named RLTrans) with heuristic-based graph transformation strategies from \cite{byilmaz_efficient_graph_trans_strategies_9eylul2026}, namely threeCriteria, 3CRI\_THICKENED, 3CRI\_AGGRESSIVE and 2CRI.}
\label{table:comparison}
\begin{tabular}{@{}lllllll@{}}
	\toprule
	\textbf{}                                                                 & \multicolumn{6}{c}{\textbf{Reduction in num. of levels}}                                                                                                                                                                                                                                                                                                                                            \\
	\multicolumn{1}{c}{\textbf{matrix}}                                       & \multicolumn{1}{c}{\textbf{\begin{tabular}[c]{@{}c@{}}num. of \\ levels\end{tabular}}} & \multicolumn{1}{c}{\textbf{\begin{tabular}[c]{@{}c@{}}three\\ Criteria\end{tabular}}} & \multicolumn{1}{c}{\textbf{\begin{tabular}[c]{@{}c@{}}3CRI\_\\ THICKENED\end{tabular}}} & \multicolumn{1}{c}{\textbf{\begin{tabular}[c]{@{}c@{}}3CRI\_\\ AGGRESSIVE\end{tabular}}} & \multicolumn{1}{c}{\textbf{2CRI}} & \multicolumn{1}{c}{\textbf{RLTrans}} \\
	\midrule
	bcsstk17                            & 1332                                           & 16\%                                       & 27\%                                         & 33\%                                          & 48\%                              & 24\%                                 \\
	bcsstk37                            & 5030                                           & 14\%                                       & 30\%                                         & 46\%                                          & 53\%                              & 24\%                                 \\
	cfd2                                & 4357                                           & 10\%                                       & 17\%                                         & 19\%                                          & 33\%                              & 3\%                                  \\
	gearbox                             & 4586                                           & 21\%                                       & 37\%                                         & 38\%                                          & 45\%                              & 16\%                                 \\
	lung2                               & 479                                            & 46\%                                       & 90\%                                         & 90\%                                          & 90\%                              & 94\%                                 \\
	PR02R                               & 2838                                           & 9\%                                        & 10\%                                         & 10\%                                          & 18\%                              & 3\%                                  \\
	torso2                              & 513                                            & 15\%                                       & 21\%                                         & 19\%                                          & 42\%                              & 18\%                                 \\
	venkat01                            & 4176                                           & 10\%                                       & 19\%                                         & 27\%                                          & 35\%                              & 4\%                                  \\
	\midrule
\textbf{\begin{tabular}[c]{@{}l@{}}avg. level \\ reduction\end{tabular}}  & \multicolumn{1}{c}{\textbf{-}}              & \textbf{18\%}                                                                         & \textbf{31\%}                                                                           & \textbf{35\%}                                                                            & \textbf{46\%}                     & \textbf{23\%}                        \\
\midrule
\multicolumn{7}{c}{\textbf{Percentage of rewritten rows}}                                                                                                                                                                                                                                                                                                                                                                                                                       \\
\multicolumn{1}{c}{\textbf{matrix}}                                       & \multicolumn{1}{c}{\textbf{\begin{tabular}[c]{@{}c@{}}num. of \\ rows\end{tabular}}}   & \multicolumn{1}{c}{\textbf{\begin{tabular}[c]{@{}c@{}}three\\ Criteria\end{tabular}}} & \multicolumn{1}{c}{\textbf{\begin{tabular}[c]{@{}c@{}}3CRI\_\\ THICKENED\end{tabular}}} & \multicolumn{1}{c}{\textbf{\begin{tabular}[c]{@{}c@{}}3CRI\_\\ AGGRESSIVE\end{tabular}}} & \multicolumn{1}{c}{\textbf{2CRI}} & \multicolumn{1}{c}{\textbf{RLTrans}} \\
	\midrule
	bcsstk17                            & 219,812                                        & 0.36\%                                     & 0.63\%                                       & 0.86\%                                        & 0.53\%                            & 1.58\%                               \\
	bcsstk37                            & 583,240                                        & 0.45\%                                     & 0.84\%                                       & 1.35\%                                        & 0.89\%                            & 0.80\%                               \\
	cfd2                                & 1,605,669                                      & 0.41\%                                     & 0.75\%                                       & 0.80\%                                        & 1.15\%                            & 0.48\%                               \\
	gearbox                             & 4,617,075                                      & 0.11\%                                     & 0.19\%                                       & 0.21\%                                        & 0.15\%                            & 0.20\%                               \\
	lung2                               & 273,647                                        & 0.17\%                                     & 0.49\%                                       & 0.49\%                                        & 0.49\%                            & 1.50\%                               \\
	PR02R                               & 4,174,236                                      & 0.18\%                                     & 0.21\%                                       & 0.22\%                                        & 0.38\%                            & 0.14\%                               \\
	torso2                              & 574,718                                        & 0.47\%                                     & 1.17\%                                       & 1.17\%                                        & 1.41\%                            & 0.98\%                               \\
	venkat01                            & 890,108                                        & 0.49\%                                     & 0.90\%                                       & 1.36\%                                        & 1.14\%                            & 0.87\%                               \\
	\midrule
	\textbf{\begin{tabular}[c]{@{}l@{}}avg. rewritten\\  rows\end{tabular}}   & \multicolumn{1}{c}{\textbf{-}}              & \textbf{0.33\%}                                                                       & \textbf{0.65\%}                                                                         & \textbf{0.81\%}                                                                          & \textbf{0.77\%}                   & \textbf{0.82\%}                      \\
	\midrule
	\multicolumn{7}{c}{\textbf{Coefficient of variation (CV) of level costs}}                                                                                                                                                                                                                                                                                                                                                                                                       \\
	\multicolumn{1}{c}{\textbf{matrix}}                                       & \multicolumn{1}{c}{\textbf{no trans.}}                                                 & \multicolumn{1}{c}{\textbf{\begin{tabular}[c]{@{}c@{}}three\\ Criteria\end{tabular}}} & \multicolumn{1}{c}{\textbf{\begin{tabular}[c]{@{}c@{}}3CRI\_\\ THICKENED\end{tabular}}} & \multicolumn{1}{c}{\textbf{\begin{tabular}[c]{@{}c@{}}3CRI\_\\ AGGRESSIVE\end{tabular}}} & \multicolumn{1}{c}{\textbf{2CRI}} & \textbf{RLTrans}                     \\
	\midrule
	bcsstk17                                                                  & 0.66                                                                                   & 0.49                                                                                  & 0.49                                                                                    & 0.63                                                                                     & 0.54                              & 0.28                                 \\
	bcsstk37                            & 0.76                                           & 0.64                                       & 0.60                                         & 0.81                                          & 0.75                              & 0.67                                 \\
	cfd2                                & 0.57                                           & 0.46                                       & 0.54                                         & 0.53                                          & 0.43                              & 0.55                                 \\
	gearbox                             & 0.88                                           & 0.68                                       & 0.67                                         & 0.57                                          & 0.75                              & 0.67                                 \\
	lung2                               & 6.72                                           & 4.89                                       & 1.86                                         & 1.86                                          & 1.86                              & 1.37                                 \\
	PR02R                               & 0.37                                           & 0.28                                       & 0.33                                         & 0.33                                          & 0.32                              & 0.33                                 \\
	torso2                              & 0.82                                           & 0.65                                       & 0.65                                         & 0.68                                          & 0.61                              & 0.46                                 \\
	venkat01                            & 0.46                                           & 0.39                                       & 0.46                                         & 0.41                                          & 0.42                              & 0.46                                 \\
	\textbf{\begin{tabular}[c]{@{}l@{}}avg. CV of\\  level cost\end{tabular}} & \textbf{1.41}                               & \textbf{1.06}                                                                         & \textbf{0.70}                                                                           & \textbf{0.73}                                                                            & \textbf{0.71}                     & \textbf{0.60}                        \\
	\botrule
	\end{tabular}
\end{table}

Table~\ref{table:comparison} compares the level reduction, coefficient of variation (CV) of level costs, and percentage of rewritten rows obtained by RL-based graph transformation(RLTrans) and the heuristic-based strategies for the best inference results obtained. In terms of level reduction, the heuristic strategies generally achieve larger reductions than RLTrans strategy. In particular, 2CRI provides the highest level reduction for most matrices, reaching $53\%$ for \textit{bcsstk37}. The RL-based transformation achieves its highest level reduction for \textit{lung2}, reducing the number of levels by $94\%$, which exceeds the $90\%$ reduction obtained by the heuristic strategies for this matrix. For the remaining matrices, RLTrans achieves reductions ranging from $3\%$ to $24\%$.

The CV results, however, provide a different perspective. RLTrans strategy achieves the lowest CV among all evaluated methods for \textit{bcsstk17} ($0.28$) and \textit{torso2} ($0.46$), while also achieving competitive CV values for \textit{gearbox} and \textit{PR02R}. In particular, the difference between level reduction and workload balance is evident for \textit{bcsstk17}: 2CRI reduces the number of levels by $48\%$, compared with $24\%$ for RLTrans, but results in a CV of $0.54$ compared with $0.28$ for RLTrans. This illustrates the trade-off between aggressively collapsing levels and maintaining a balanced distribution of level costs. The RL-based approach does not consistently outperform the heuristic strategies in either objective individually; rather, its learned transformations provide a different balance between level reduction and workload distribution. For \textit{lung2}, for example, RL simultaneously achieves the highest level reduction ($94\%$) and a substantial CV reduction from $6.72$ to $1.37$.

The level reduction and CV results should also be considered together with the amount of graph rewriting required to obtain these transformations. RLTrans rewrites between $0.14\%$ and $1.58\%$ of the rows across the evaluated matrices. The results demonstrate that meaningful changes in both level structure and workload distribution can be obtained by modifying only a small fraction of the rows. For \textit{bcsstk17}, RL rewrites more rows and achieves a lower level reduction than the heuristic approaches, but obtains a substantially better CV. For \textit{lung2}, RLTrans rewrites more rows while achieving both a higher level reduction and a better CV than the heuristic approaches. For \textit{torso2}, RLTrans rewrites fewer rows while achieving a level reduction comparable to the heuristic approaches and a lower CV. Finally, for \textit{bcsstk37}, the level reduction, CV, and percentage of rewritten rows obtained by RLTrans are comparable to those of the heuristic approaches.

Among the graph transformation strategies presented in the table, RLTrans involves the largest fraction of rewritten rows. However, it should be noted that the percentage of rewritten rows does not directly represent the number of rewriting operations for the RL-based approach. Unlike the heuristic strategies, where a row is rewritten from its source level to a target level only once, the RLTrans agent transforms the graph sequentially and may select a previously rewritten row again as the graph state evolves. Consequently, the reported percentage represents the fraction of rows involved in the learned transformation rather than a direct count of rewrite operations. 

Overall, the comparison demonstrates that the RL-based approach and the heuristic strategies exhibit different optimization characteristics. The heuristic strategies, particularly 2CRI, are generally more effective at aggressively reducing the number of levels, whereas RLTrans can provide stronger improvements in the balance of level costs for some matrices. This difference is consistent with the multi-objective nature of the transformation problem: collapsing levels can increase the workload of the remaining levels, while improving the balance of level costs may limit the achievable level reduction. The RLTrans agent addresses this trade-off through sequential, state-dependent transformations rather than a fixed heuristic transformation sequence.

\section{Conclusion and Future work} \label{section:conclusion}
This work presents a reinforcement learning-guided graph transformation framework for sparse triangular solves within the ChainBreaker framework. The proposed approach formulates graph transformation as a reinforcement learning problem while using a deterministic graph rewriting engine to apply graph transformations. Unlike heuristic-based approaches, which require manually designed transformation rules, the proposed framework enables the automated discovery of graph transformation policies through interaction with the graph transformation environment by providing a flexible experimental platform for developing and evaluating future graph transformation strategies. It enables the systematic investigation of graph representations, reward formulations, and transformation policies without modifying the underlying deterministic graph rewriting engine.

The developed environment models graph transformation using matrix-derived features that capture workload balance and parallelism, allowing the reinforcement learning agent to optimize graph transformations for improved SpTRSV performance. 

Experimental results on real-world sparse matrices demonstrate that the RL-based approach can learn effective matrix-dependent transformation policies. Across the evaluated matrices, RLTrans achieves an average level reduction of $23\%$ and an average reduction of $29\%$ in the coefficient of variation of level costs, with individual level reductions reaching up to $94\%$ and CV reductions reaching up to $80\%$. The comparison with heuristic strategies further demonstrates the trade-off between level reduction and workload balance. The heuristic strategies generally achieve more aggressive level reduction, whereas RLTrans achieves a larger average reduction in CV of level costs of $29\%$. This behavior demonstrates that the learned policy can navigate competing graph transformation objectives.

The amount of graph modification required by the RLTrans approach is also relatively small. Across the evaluated matrices, between $0.14\%$ and $1.58\%$ of the rows are involved in the learned transformations, with an average of only $0.82\%$. Despite modifying such a small fraction of the rows, RLTrans produces substantial changes in both the level structure and the distribution of level workloads. These results demonstrate the potential of reinforcement learning as an automated and matrix-dependent approach to graph transformation.

The transfer learning experiments further demonstrate that graph transformation policies can benefit from previously acquired knowledge. Curriculum learning followed by fine-tuning provides useful transfer to previously unseen matrices, although the quality of the transferred policy varies with the characteristics of the target matrix. In contrast, zero-shot inference generally provides weaker transformation results, which is consistent with the strong dependence of graph transformation behavior on the underlying sparsity pattern. Nevertheless, the zero-shot experiments provide evidence that some learned transformation behavior can be applied to previously unseen matrices without additional training and provide a useful basis for studying generalization.

Immediate future work will focus on collecting SPTRSV performance results on solvers from \cite{byilmaz_ALB_2020_HPCAsia} and \cite{byilmaz_efficient_graph_trans_strategies_9eylul2026} as well as an SpTRSV library implementation, to evaluate the applicability of the proposed graph transformations across different solver implementations. Training will also be extended to a broader collection of sparse matrices. These steps will prepare the framework for further research directions including incorporating hardware-aware transformation objectives, such as memory access locality and architecture-specific optimization criteria, as well as developing policies specialized for different sparse triangular solve implementations.

\section{Related Work}
\label{section:related_work}
Several parallelization strategies for sparse triangular solve (SpTRSV) have been proposed in the literature. Existing approaches can be broadly classified into level-set methods~\cite{AndersonS89_Saad,Li2013,naumov_2011,rothberg_1992_parallel_iccg,Saltz:1990:aggregation} and synchronization-free methods~\cite{aliaga201979,hammond_schreiber_1992,Li_2017,liu_sync_free,liu_2017_fast_synchronization_free_algorithms}. Both categories exploit the directed acyclic graph (DAG) induced by the sparse triangular matrix to expose parallelism while respecting data dependencies among rows.

The level-set method partitions the DAG into a sequence of levels such that all rows assigned to the same level are mutually independent and can therefore be processed concurrently. Rows in a level depend only on rows belonging to preceding levels, requiring each level to complete before the next one begins. Consequently, synchronization barriers are introduced between consecutive levels.
 
 The effectiveness of the level-set method is strongly influenced by the sparsity pattern of the input matrix, which determines both the number of levels and the computational workload assigned to each level. Matrices containing many sparsely populated levels or exhibiting substantial workload imbalance across levels often experience poor parallel efficiency. Thin levels underutilize the available processing resources, while significant differences in computational cost between levels lead to load imbalance. Moreover, each level transition introduces a synchronization barrier, increasing waiting time as threads must remain idle until all computations in the current level have completed. As a result, the overall performance of level-set methods is highly sensitive to both the critical path length and the distribution of workload across levels. Early implementations of the level-set method primarily targeted multicore CPU architectures~\cite{AndersonS89_Saad,rothberg_1992_parallel_iccg,Saltz:1990:aggregation}. With the growing adoption of many-core accelerators, the same execution model has subsequently been adapted for GPUs~\cite{Li2013,naumov_2011}.

Synchronization-free methods emphasize load balancing by partitioning the rows into groups that are assigned to processing units. Unlike the level-set method, rows within the same group may have dependencies among themselves, providing greater flexibility when distributing the workload. Computation begins as soon as the dependencies of a group are satisfied, allowing synchronization to occur at a finer granularity. Consequently, synchronization-free methods rely on lightweight synchronization mechanisms, such as locks, rather than global barriers. The size of the row groups can be adapted to the characteristics of the target architecture. However, maintaining synchronization metadata for all groups and the use of spinning locks may introduce considerable overhead for large sparse matrices. Successful implementations of synchronization-free methods have been reported for both multicore CPUs~\cite{hammond_schreiber_1992} and GPUs~\cite{aliaga201979,Li_2017,liu_sync_free,liu_2017_fast_synchronization_free_algorithms,su2020}.

In general, level-set methods are particularly well suited to multicore CPU architectures, where a moderate number of threads execute rows within a level concurrently before synchronizing at barrier points. Synchronization-free methods, on the other hand, have shown particular promise on GPU architectures, whose massive thread parallelism and efficient hardware scheduling make fine-grained synchronization more practical.

To combine the advantages of both execution models, hybrid approaches integrating level-set and synchronization-free techniques have also been proposed~\cite{byilmaz_ALB_2020_HPCAsia,Park:2014:SSH:2769884.2769893}. These approaches seek to reduce synchronization overhead while preserving sufficient parallelism by eliminating unnecessary dependencies in the dependency graph.

Several other optimization techniques have also been proposed for SpTRSV. Graph-coloring methods expose parallelism by partitioning the dependency graph into independent color sets and have been employed as a load-balancing strategy on both multicore CPUs~\cite{iwashita_2012,Ma1998DistributedIA} and GPUs~\cite{naumov_2015,Suchoski_2012}. 

Blocking techniques constitute another widely adopted optimization strategy. These methods partition the sparse triangular matrix into diagonal triangular blocks and off-diagonal rectangular blocks, creating denser computational regions that improve cache locality and data reuse. Various partitioning strategies have been investigated in the literature. For example, Lu et al.~\cite{lu2020} proposed column-block, row-block, and iterative-block algorithms for GPUs and selected the most suitable blocking strategy according to the sparsity characteristics of the input matrix. Since blocking typically requires matrix reordering, these approaches incur additional preprocessing overhead, and their effectiveness largely depends on the sparsity structure of the input matrix. Several blocking-based optimizations have been proposed for both CPU and GPU architectures~\cite{mayer2009,Smith:2011:STS:2076556.2076558,totoni_2014_structure_adaptive,CUGU2020371,lu2020,su2020,zhang2021}.

Compiler-assisted optimization represents another important research direction for sparse matrix computations. By exploiting domain-specific knowledge, compilers can perform context-aware transformations that are difficult to achieve with general-purpose optimization passes. Examples include domain-specific code generation frameworks such as Sparso~\cite{rong_park_sparso_2016} and Sympiler~\cite{Cheshmi:2017:Sympiler}, which apply optimizations including dense-block extraction, loop transformations, elimination of indirect memory accesses for sparse computations, arithmetic simplifications, and memory-access optimizations~\cite{Cheshmi:2017:Sympiler}. AG-SpTRSV~\cite{hu2024} introduced a GPU framework that generates multiple code variants using a performance model and performs workload balancing by assigning groups of rows to GPU warps according to matrix characteristics such as block size and average row density. More recently, Patel et al.~\cite{patel2025} proposed a compiler capable of generating symmetry-aware code for sparse and structured tensor computations.

\section*{Declarations}

This work was supported by T\"{u}rkiye Bilimsel ve Teknolojik Ara\c{s}t{\i}rma Kurumu (T\"{U}B\.{I}TAK) with  Grant/AwardNumber: 121E612.

The authors declare that they have no competing interests or conflicts of interest related to this work.

\section*{Availability of Data and Materials:}
The code developed for this work is available in the github repository mentioned on page 15 footnote. In addition, all data generated or analysed during this study are included in this published article.

Data availability: The datasets generated and analyzed during the current study are available from the corresponding author on reasonable request since it is not put in a publicly avaiable repository yet.


%
%
%
%
%
%
\begin{appendices}

\section{Level Cost Change by RL Transformation}\label{secA1}
 \begin{figure}[!htbp]
	\centering
	\includegraphics[width=0.9\textwidth]{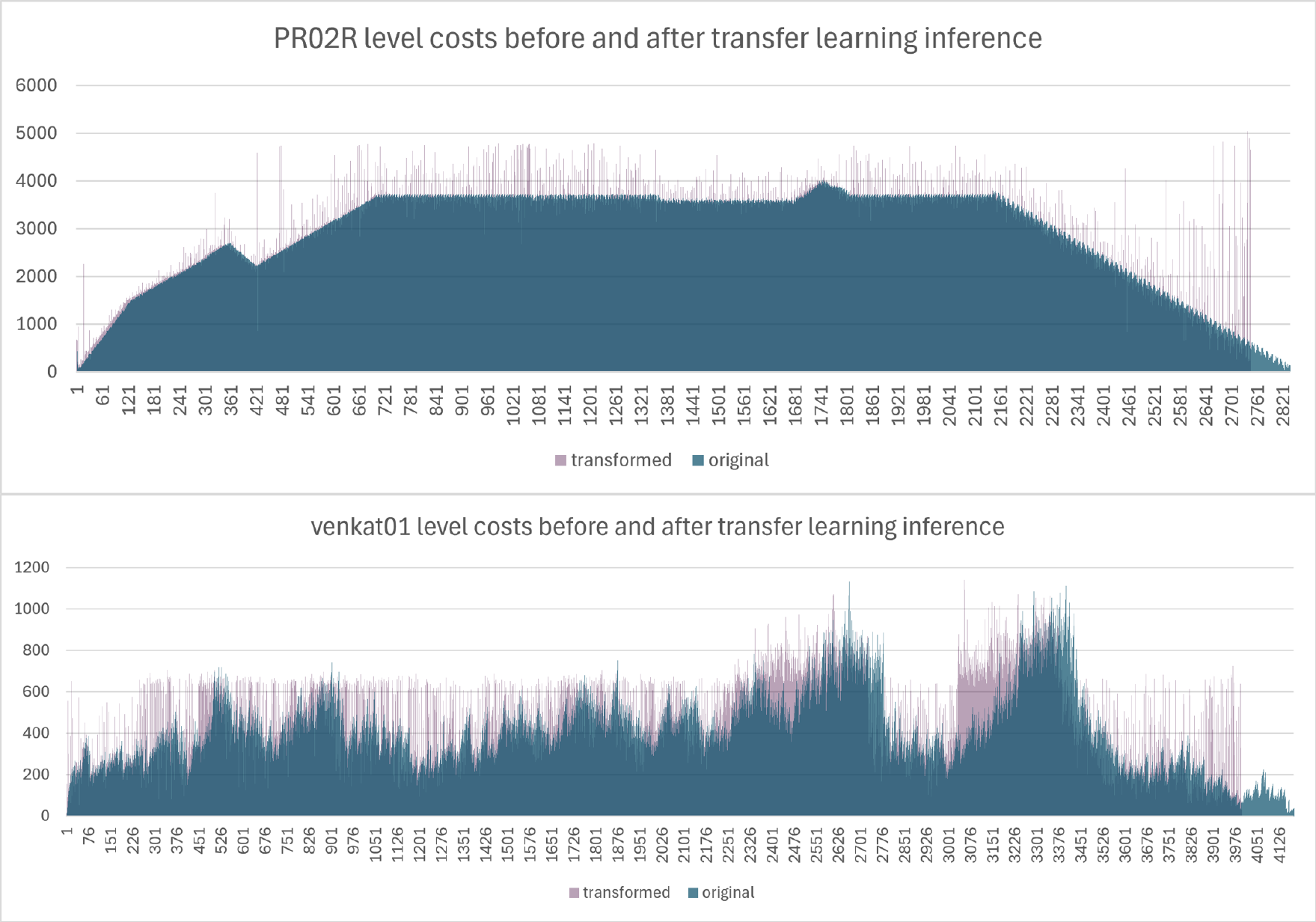}
	\caption{Level costs before and after RL graph transformation.}\label{fig:group2}
\end{figure}
 \begin{figure}[!htbp]
	\centering
	\includegraphics[width=0.85\textwidth]{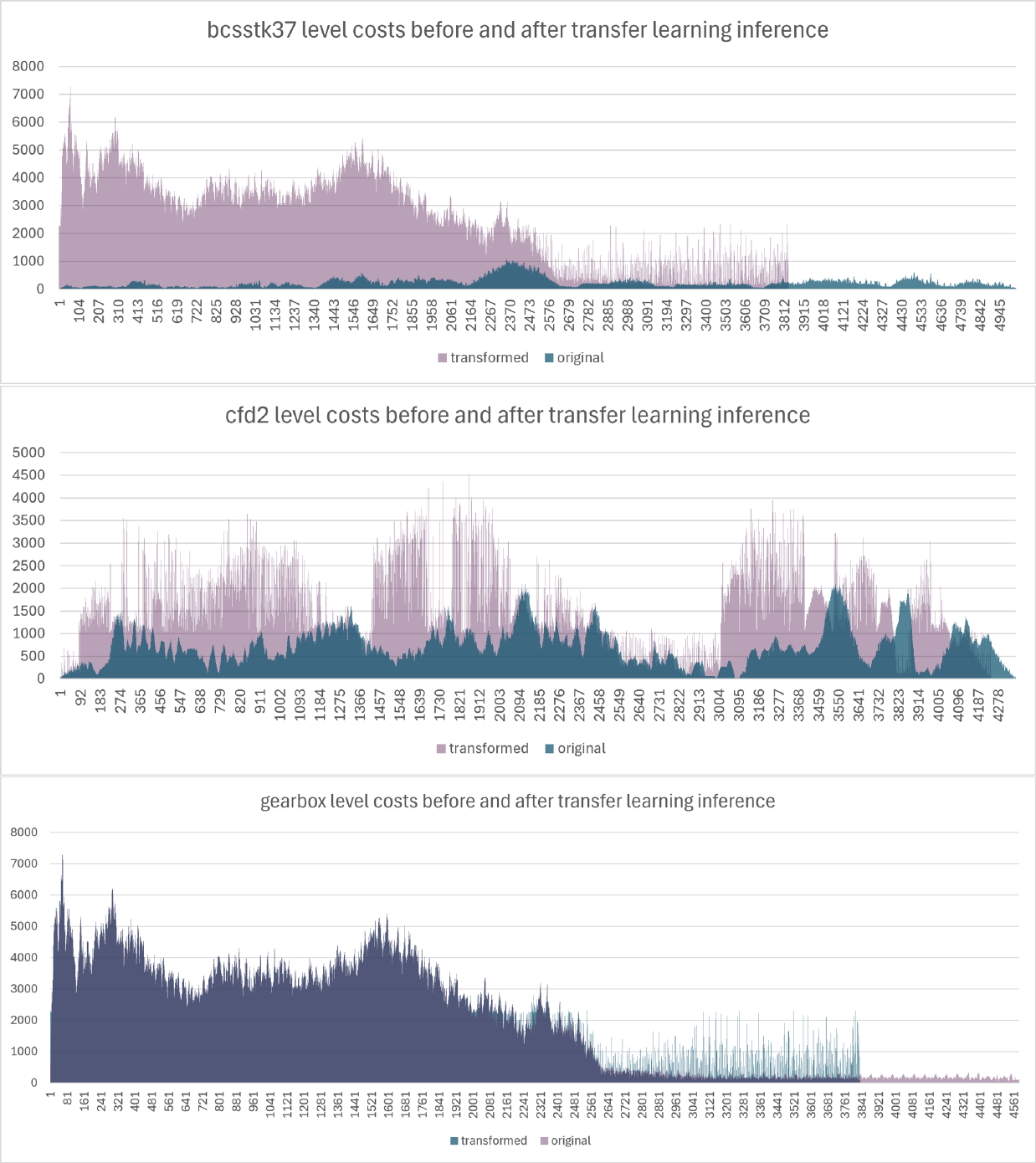}
	\caption{Level costs before and after RL graph transformation.}\label{fig:group1}
\end{figure}





\end{appendices}


\bibliography{sn-bibliography}

\end{document}